\documentclass[10pt,conference]{IEEEtran}

\IEEEoverridecommandlockouts

\usepackage{amsmath,amssymb,amsfonts}

\usepackage{algorithmic}

\usepackage{cite}

\usepackage{graphicx}

\usepackage{textcomp}

\usepackage{xcolor}

\def\BibTeX{{\rm B\kern-.05em{\sc i\kern-.025em b}\kern-.08em

 T\kern-.1667em\lower.7ex\hbox{E}\kern-.125emX}}

\usepackage{xspace}

\usepackage{url}
\usepackage{xurl} 
\usepackage{hyperref}

\hypersetup{
    hidelinks,         
    breaklinks=true    
}

\newcommand{\storage}{{\cal S}}
\usepackage[mode=buildnew]{standalone}
 \usepackage{booktabs}

 \usepackage{cleveref}

 \usepackage{subcaption}

\usepackage{multirow}

\usepackage{array}

\usepackage{xspace}

\newcommand{\zstd}{\texttt{zstd}\xspace}

\newcommand{\zlib}{\texttt{zlib}\xspace}

\newcommand{\zstdlvl}[1]{\texttt{zstd-#1}\xspace}

\newcommand{\boffa}{Boffa {\em et al.}\xspace}
\newcommand{\mbs}{MiB/s\xspace}

\newcommand{\gbs}{GiB/s\xspace}

\newcommand{\kraken}{Kraken\xspace}
\newcommand{\grenoble}{Grenoble Alpes Recherche/GRICAD\xspace}
\newcommand{\bettik}{Bettik\footnote{\url{https://gricad-doc.univ-grenoble-alpes.fr/hpc/data_management/bettik}}\xspace}
\newcommand{\hoyt}{Hoyt\footnote{\url{https://gricad-doc.univ-grenoble-alpes.fr/hpc/data_management/silenus}}\xspace}

\begin{document}
\title{The Energy--Throughput Trade-off in Lossless-Compressed Source Code Storage\\

\thanks{This study was funded by the Alfred P. Sloan Foundation with the grant \#\href{https://sloan.org/grant-detail/g-2025-25193}{G-2025-25193} (\url{sloan.org}).}
\thanks{© 2026 IEEE. Personal use of this material is permitted. Permission from IEEE must be obtained for all other uses, in any current or future media, including reprinting/republishing this material for advertising or promotional purposes, creating new collective works, for resale or redistribution to servers or lists, or reuse of any copyrighted component of this work in other works.}
\thanks{The Version of Record of this contribution will be published in the Proceedings of the SANER 2026 - Greenvolve Workshop, and is available online at: \url{https://doi.org/[TBA]}}

}

\author{\IEEEauthorblockN{Paolo Ferragina}
\IEEEauthorblockA{\textit{L'EMbeDS Department} \\
\textit{Sant'Anna School of Advanced Studies}\\
Pisa, Italy \\
\href{mailto:Paolo.Ferragina@santannapisa.it}{Paolo.Ferragina@santannapisa.it}
}

\and

\IEEEauthorblockN{Francesco Tosoni, \textit{Member, IEEE}}
\IEEEauthorblockA{\textit{L'EMbeDS Department} \\
\textit{Sant'Anna School of Advanced Studies}\\
Pisa, Italy \\
\href{mailto:Francesco.Tosoni@santannapisa.it}{Francesco.Tosoni@santannapisa.it}
}
}
\maketitle

\begin{abstract}

Retrieving data from large-scale source code archives is vital for AI training, neural-based software analysis, and information retrieval, to cite a few. This paper studies and experiments with the design of a compressed key-value store for the indexing of large-scale source code datasets, evaluating its trade-off among three primary computational resources: (compressed) space occupancy, time, and energy efficiency. 
Extensive experiments on a national high-performance computing infrastructure demonstrate that different compression configurations yield distinct trade-offs, with high compression ratios and order-of-magnitude gains in retrieval throughput and energy efficiency. We also study data parallelism and show that, while it significantly improves speed, scaling energy efficiency is more difficult, reflecting the known non-energy-proportionality of modern hardware and challenging the assumption of a direct time--energy correlation. 
This work streamlines automation in energy-aware configuration tuning and standardized green benchmarking deployable in CI/CD pipelines, thus empowering system architects with a spectrum of Pareto-optimal energy--compression--throughput trade-offs and actionable guidelines for building sustainable, efficient storage backends for massive open-source code archival.

\end{abstract}

\section{Introduction}

The proliferation of Large Language Models (LLMs) such as Gemini, ChatGPT, Claude, and the like~\cite{llm_writing, llm_zenodo} has exacerbated the competition for fetching human-written digital texts~\cite{mozilla2025bestpractices, llama-open}. The resulting large-scale {\em scraping} strains open archives, like Wikimedia Commons~\cite{Mueller2025} and Software Heritage (SWH)~\cite{di-cosmo-archiving, zacchiroli-archiving-repro, codecommons}. Thus, Wikimedia's reported 65\% bot traffic with 50\% bandwidth growth in 2025~\cite{Mueller2025} has unfortunately limited bandwidth for traditional human access. The Stack v1 and v2~\cite{the-stack, the-stack-v2} and related SWH-derived source code datasets, essential for training code-generation models like GitHub Copilot~\cite{copilot-substitute-human, copilot-robustness} or Code Llama~\cite{llama-release}, are receiving thousands of downloads requests per month.
It is no surprise that the SWH Archive eventually resolved to restrict LLM bulk fetching\footnote{\url{https://www.softwareheritage.org/legal/bulk-access-terms-of-use}}. Yet, SWH's Winery backend, with its block-wise (de)compression on unsorted Ceph data~\cite{ceph}, is still too slow. Even Wikimedia imposed similar restrictions: a maximum of 2 robotic connections at 25 Mbps\footnote{\url{https://wikitech.wikimedia.org/wiki/Special:PermaLink/2324939}}. Yet, all these approaches, alongside degrading file quality~\cite{degrading_data} or paid services~\cite{cacm_quo_vadis}, impair libre access and tackle symptoms, not the structural mismatch between traditional infrastructure and current demands in the LLM era. Recent industrial initiatives and EU regulations thus targeted bandwidth limits in archival and advocated for funding software development~\cite{dicosmo-nature-stop}.
There is an energy concern, too. Energy-hungry datacenters consume 2\% of global electricity, yet programmers still neglect energy efficiency~\cite{programmers-know-energy}: storage's embodied footprint~\cite{dirty_SSD} worsens as AI-driven disk demand increases. Hence, archive sustainability too covers a key innovation role~\cite{ailamaki2025cambridgereportdatabaseresearch} and (green) software engineers combine modeling and eco-conscious techniques to entail energy-related evaluations. Addressing these structural and sustainability challenges requires automated, energy-aware systems that can be seamlessly integrated into modern CI/CD pipelines.
Driven by scalability concerns with SWH, we design a key-value store for compressed source code and evaluate its space-time-energy performance. Our work extends static compressed formats for source code by \cite{BoffaSWH}, limited to \mbs retrieval. Our system accommodates all their datasets in a single instance, thus scaling up to a Terabyte-scale indexed codebase while providing a dynamic, energy-aware system that consistently achieves \gbs retrieval throughput with similar compressed size. 

\begin{figure}

  \centering

  \includegraphics[width=.85\linewidth]{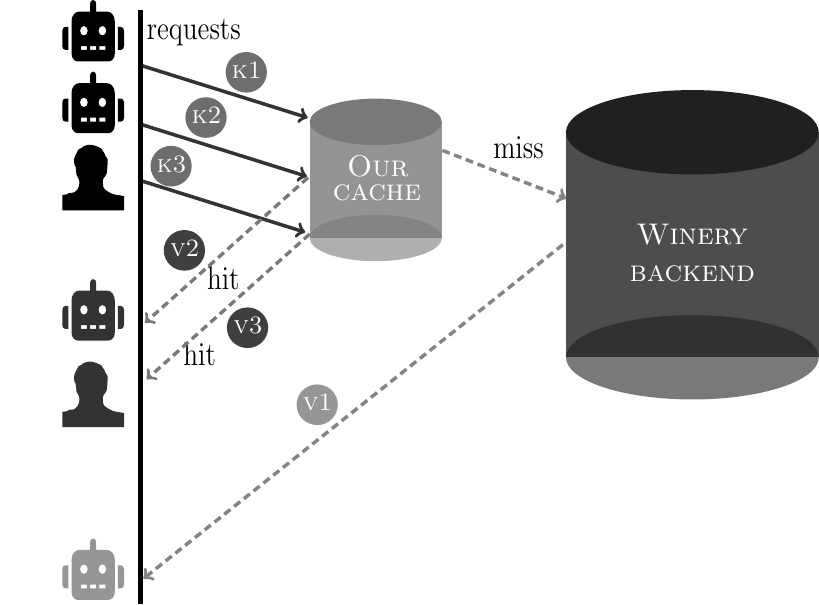}

  \caption{A multi-tier storage architecture, which first directs key lookups to our PPC-based cache; it forwards queries to the Winery backend only upon a cache miss.}\label{fig:multitier}

\end{figure}

Our experiments on a national HPC infrastructure show that there exist configurations achieving significant compression, with order-of-magnitude gains in throughput and energy efficiency. We note that data parallelism makes completion time scale better than energy efficiency, reflecting hardware non-energy-proportionality. As a result, our findings equip system architects with a spectrum of Pareto-optimal configurations and practices for building greener, efficient storage backends for massive open-source code archives, and beyond.

\section{Problem Definition and Results}

\label{sec:problem}

Let us formalize the research scenario. We are given a stream of operations $\{q_1, \ldots, q_m\}$, where each $q_i$ is an insert, lookup, update, or removal over a source-code collection organized in a two-level storage scheme: a local disk of size $M$, and a large disk-based storage $\storage$. Our goal is then to design a time--energy performant compressed cache of size $\leq M$ supporting these operations with minimum access to $\storage$. The cache thus indexes a well-selected subset of frequently-used SWH files, acting as the last in a multi-tier object storage system (as in \Cref{fig:multitier}) backed by a larger yet slower store, such as SWH's Ceph-based Winery backend\footnote{\url{https://docs.softwareheritage.org/devel/swh-objstorage/winery.html}}~\cite{ceph}, hosting the exhaustive archive. This way, our customized cache can tailor to geographical request patterns or scraping bots, or provide a data pump for language- and code-to-code retrieval. We establish sortable indexing keys for a lossless-compressed NoSQL key-value store. This key design and the algorithmic idea of implementing the Permute-Partition-Compress (PPC) paradigm (see \Cref{subsec:ppc}) via a key-value store is our cornerstone for boosting compression by exploiting inter-file redundancies and achieving dynamicity and scalability at the Terabyte scale. 

We proceed by exploring three main research questions:

\begin{enumerate}

    \item[Q1]How can a commodity key-value store implement the PPC paradigm to overcome the scale and static limitations of \cite{BoffaSWH}, therefore achieving robust time, space, and energy performance at the Terabyte scale?

  \item[Q2] What is the interplay between data size, compressibility, and the time--energy trade-off in our proposed solution?

  \item[Q3] For which query settings and parallelism configurations can our solution achieve a large-scale and energy-aware retrieval of source-code files?

\end{enumerate}

\noindent Our architectural goal is to introduce a SWH multi-tier storage engine, as depicted in \Cref{fig:multitier}, with a large-scale cache based on a principled combination of the PPC paradigm and a commodity key-value store. This combination allows for different space--time Pareto solutions for specific objectives. Though our solution is fully dynamic, herein we limit our considerations to building (which provides a proxy for batched or burst insertion) and the retrieval phase; we do not test deletions. Related to the three questions above, for Q1 we will design and analyze the impact of our proposal across multifaceted space-, time-, and energy-diverse configurations, observing compelling trade-offs in time-energy retrieval, construction efficiency, compression ratios (defined as the ratio between compressed and uncompressed data size: the lower the better), and thread count. In particular, there are proper configurations of our key-value store design that enable up to $3\times$ time--energy build cost reduction and consistently improve time--energy retrieval performance by one to two orders of magnitude. For Q2, we will show that stronger compression reduces space but sacrifices build time--energy performance. Retrieval performance is nuanced, in that a complex interplay exists amongst block size, compression level, and thread count: a balanced solution using \zstdlvl{6} with a small block of 4 KiB dominates retrieval performance (in the order of at least a few GB/s), though with scalability limited to 16-32 threads. For Q3, we confirmed that the substantial scalability of our solution for uniformly sampled key queries (8.4-30.0$\times$ for retrieval) exhibits no single, unique optimum; instead, a rich Pareto-optimal spectrum exists, enabling architects to adapt our and similar green, source-code storage systems to various workloads. We also consider a skewed query distribution, reporting greater retrieval throughput at lower parallelism degrees. Our methodology and outcomes are nonetheless adequately general for extension to any redundant text-based collection and key-value storage architecture.

We introduce in \Cref{sec:basic} notions and background, before moving to the design of our compressed key-value store (in \Cref{sec:proposal}). \Cref{sec:setup} delves into the code setup, entailing HPC infrastructure (\ref{subsec:hpc_infra}), data features (\ref{subsec:datasets}), and RocksDB tuning (\ref{sec:config}), the key-value store we adopted. In heading toward the end, we report in \Cref{sec:micro-bench} on micro-benchmarks for a 10-GiB Python corpus and evaluate, in \Cref{sec:scaling}, large-scale multi-criteria performances. To conclude, \Cref{sec:conclusions} sums up our findings and future work.

\section{Basic Concepts}
\label{sec:basic}

\subsection{Software Heritage}
\label{subsec:swh}

\begin{figure}

  \centering

  \includegraphics[width=.99\linewidth]{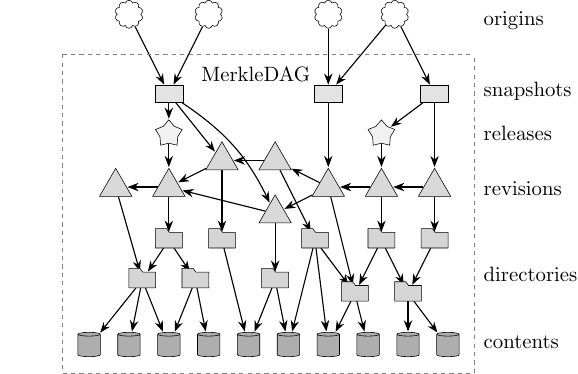}

  \caption{Software Heritage data model with source contents representing the leaves of the DAG}\label{fig:swh-dag}

\end{figure}

Paper \cite{what-preserve2006} considers source code a full-fledged scientific and industrial human artifact, which deserves preservation. It is this awareness that gives ground to the Software Heritage (SWH) initiative and infrastructure, backed by Inria and UNESCO, to gather, curate, and maintain the world's largest open-source code archive~\cite{di-cosmo-archiving, zacchiroli-archiving-repro}. SWH enables tracing software evolution~\cite{SWH-filepath-assignment}, to monitor vulnerabilities~\cite{SWH-cybersecurity}, and provides libre data for AI training. UNESCO supports SWH, because it aligns with UN-led agendas for equitable AI and for open science\footnote{\url{https://unesdoc.unesco.org/ark:/48223/pf0000383771}}. The CodeCommons initiative contributes to SWH goals by carefully selecting code subsets for distribution in suitable Parquet formats and AI applications, as done for code datasets like BigCode's The Stack~\cite{the-stack, the-stack-v2}. It has been observed~\cite{few-shot-commonsense} that such training code sets potentiate LLMs' reasoning, even general-purpose ones. Year after year, SWH has been ingesting from GitHub, GitLab, and many more origins, archiving more than 20 billion files and over 2 Petabytes from over 400 million projects (October 2025)~\cite{SWHAP_conf}\footnote{\url{https://unesdoc.unesco.org/ark:/48223/pf0000371017}}. These massive data impose infrastructural challenges not just for storage, but for indexing and retrieval too. Code (meta)data form a Directed Acyclic Graph (DAG) in \Cref{fig:swh-dag} with contents alone (``blobs'') consuming 99\% of storage even after deduplication and compression~\cite{dicosmo_why_how}. Lowering the current SWH Filesystem (SwhFS)\footnote{\url{https://docs.softwareheritage.org/devel/swh-fuse}} storage costs is hence crucial.
\subsection{Permute, Partition, and Compress}

\label{subsec:ppc}
We base our approach on the Permute, Partition, Compress (PPC) paradigm of \cite{Buchsbaum2000, Buchsbaum2003, textual-web}, a versatile strategy to enhance compression by grouping together similar data. Though widely applicable, to our knowledge, the only prior attempt at a PPC-informed compressed format for source code is the one by \cite{BoffaSWH}, whose experiments were nonetheless limited to static code collections and to 200 GiB in size, offered modest compression and retrieval speeds (often $\approx$10 MB/s), were single-threaded, and lacked support for dynamic updates.

We extend this line of research to large-scale collections by proposing a novel implementation of the PPC paradigm hinging on a key-value store, namely RocksDB~\cite{rocksdb-evo}, which we chose for its multi-threaded performance and scalability. RocksDB's Log-Structured Merge-trees natively organize data into compressed blocks of sorted keys, making it a natural fit for the PPC paradigm and offering the additional advantage of dynamic operations over indexed files, unlike the approach of \cite{BoffaSWH}. Our goal is thereby space--time, sustainable cache for the dynamic source-code caching problem in \Cref{sec:problem}, which can effectively exploit PPC-informed key design to reduce by roughly one order of magnitude the original data size while providing efficient build and lookup operations. The following section provides additional algorithmic insights into our proposal.

\section{Our Proposal}
\label{sec:proposal}

We propose a PPC cache acting as the last in a multi-tier storage system like the one in \Cref{fig:multitier}. Our cache indexes a personalized codebase tailored to user- or application-specific needs, informed by geographical regions, the demands of data-scraping bots, and search engines within the CodeCommons project \cite{codecommons}. We assume that less frequently accessed data are held in a larger, slower object storage system, like SWH's Ceph-based Winery backend\footnote{\url{https://docs.softwareheritage.org/devel/swh-objstorage/winery.html}}~\cite{ceph}.
We thus tackle the problem of creating a fast, dynamic, large-scale PPC-based associative storage based on RocksDB \cite{rocksdb-evo}. We chose RocksDB for $\storage$ due to its open-source nature, production-grade robustness, thorough literature review~\cite[\S3]{kleppmann_book}, parallel read support, and proven scalability. It supports all dynamic operations, including updates (implemented as idempotent insertion) and deletions. Though we focus only on batched insertions for construction and retrieval of source-code files, our design accommodates dynamic updates and deletions too, instrumental for SWH's dynamic nature. Our solution also enforces a size constraint $M$ on the cache, in the Terabyte order. 
Defining the {\em key} that indexes each {\em value} (the source code) is the most critical design choice within the PPC pipeline implemented via a key-value store. A successful key assignment shall map similar file contents (values) to lexicographically-proximate keys, enabling efficient compression via ``linearized grouping''. \cite{BoffaSWH} investigated context- and content-based key designs, concluding that filenames offer the best trade-off between compression ratios and computational speed. We thus implement the key by deploying an assigned filename, derived as follows. Given the multiple filenames in the SWH archive's Merkle tree associated with each piece of content, the method described in \cite{SWH-filepath-assignment} uses a ``Popular Content Filenames Dataset'' to determine the most frequent filename as the canonical identifier. We then construct the filename-based key by prepending the file extension to the filename (e.g., {\tt h.doxygen}), followed by a unique SWH-identifier (so-called SWHID\footnote{\url{https://www.swhid.org}}) to ensure each key is distinct. We thus group files first by extension (as a language proxy), then by filename, ensuring similar files are adjacent in RocksDB serialization layout and compression blocks.

\begin{figure*}

  \centering

  \includegraphics[width=.95\textwidth]{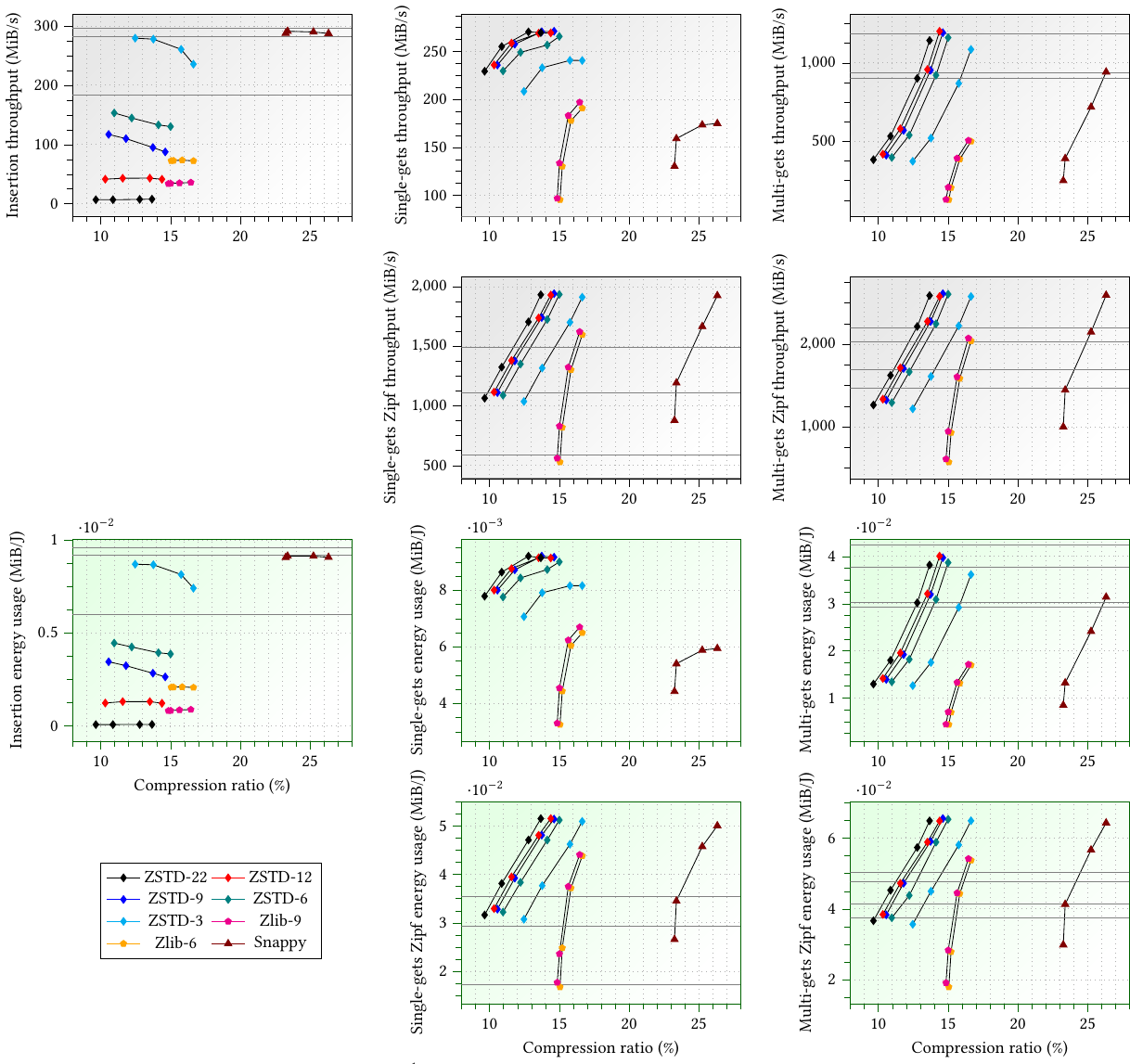}

    \caption{Our micro-benchmarks on 10-GiB Python code. The $y$-axis reports throughput (\mbs for time, MB/J for energy); the $x$-axis shows the compression ratio. The best system configurations are in the top-left corner of each plot.}\label{fig:benchmark}

\end{figure*}

\section{Code Setup}
\label{sec:setup}

\subsection{High-Performance Computing Infrastructure}
\label{subsec:hpc_infra}

We ran all experiments on the CPU partition of the \kraken{} cluster, a French national High-Performance Computing and Data Analysis (HPCDA) infrastructure operated by \grenoble{}. We leveraged the CPU cluster for multicore concurrency in our C++/CMake parallel stream code, implemented via standard POSIX threads (aka pthreads).
The CPU partition comprises several node types, all equipped with the same processors (2$\times$ AMD EPYC 9654 96C 2.4 GHz, totaling 192 cores), local storage (1.92 TB NVMe SSD), and a 200 Gb/s InfiniBand NDR HCA interconnect. We executed our cache on the 44 standard nodes with 768 GB of RAM. The OAR v3 batch scheduler handled cluster workload management. To guarantee consistent and reproducible environment conditions, we systematically restricted launched jobs to standard nodes via {\tt -p "fat = 'NO'"}.
We stored the input on \bettik{}, a high-performance shared scratch store, a BeeGFS-based filesystem with low-latency access and sustained transfer rates of up to 10 Gb/s. We stored all subsequent RocksDB serializations on \hoyt{}, a high-performance distributed scratch system built on all-SSD infrastructure with NVMe controllers. Leveraging BeeGFS and connected via a 200 Gb/s InfiniBand network with RDMA support, Hoyt delivers low-latency, high-throughput I/O, suitable for I/O-intensive and parallel workloads.

Given the impracticality of connecting power meters to each cluster node, we used the Linux profiler {\tt perf} for energy estimation. On our AMD infrastructure, we rely on the system's ability to report energy data through its performance counters, a functionality analogous to Intel's RAPL interface, which operates via Model-Specific Registers (MSRs)~\cite{rapl-in-action}. We measured at the package level, which aims to capture the power consumption of the entire processor socket, showing how software engineers can leverage readily available profiling tools to make energy-aware decisions without specialized hardware. RAPL-like interfaces are known for margins of error, yet we contend relative configuration comparisons remain valid, with adoption of {\tt Perf}, e.g., in \cite{ieee-access-2025, rapl-in-action}.

\subsection{Datasets}

\label{subsec:datasets}
For our experiments, all input data adopted the Parquet columnar format, compressed via Snappy, which is the default in libraries like Pandas and PyArrow.
\begin{table}[ht]

\centering

\begin{tabular}{lrrrc}

\toprule

\textbf{Language} & \textbf{Size (GiB)} & \textbf{Num. files} & \multicolumn{2}{r}{\textbf{File size (KiB)}} \\

\cmidrule(lr){4-5}

& & & \textbf{Mean} & \textbf{Median} \\

\midrule

Python & 200.0 & 9.640.731 & 21.7 & 7.2 \\

C/C++ & 200.0 & 6.437.613 & 32.5 & 9.0 \\

JavaScript & 200.0 & 3.464.374 & 60.5 & 4.6 \\

Java & 200.0 & 26.373.974 & 7.9 & 2.4 \\

\bottomrule

\end{tabular}

\caption{Features of the dataset by \boffa{}~\cite{BoffaSWH}.\label{tab:boffa_dataset}}

\end{table}

We considered source code from four major languages: Python, Java, JavaScript, and C/C++. The original 200-GiB tar files were converted to Parquet and divided into row groups of 100,000 rows to balance I/O efficiency and memory usage. Larger row groups improve compression and reduce metadata overhead, but increase memory demands. This collection's features, previously used by \cite{BoffaSWH} to analyze intra- and inter-file similarity within SWH, are detailed in \Cref{tab:boffa_dataset}. Before indexing large data (in \Cref{sec:scaling}), we conducted microbenchmarks (in \Cref{sec:micro-bench}) on 434,322 Python files (10 GiB). We selected Python because \cite{BoffaSWH} showed it is highly compressible, providing ideal conditions for our compressed index capabilities and enabling direct comparison with prior work.

\subsection{RocksDB Configuration}

\label{sec:config}

For reproducibility, we configured RocksDB for our hardware and write-once/read-many workload, leveraging 768 GiB of node RAM to improve time, space, and energy use. Key adjustments from defaults included enabling memory-mapped I/O ({\tt allow\_mmap\_reads}, {\tt allow\_mmap\_writes}), enlarging write buffers ({\tt write\_buffer\_size} set to 2 GiB), and optimizing for concurrency ({\tt use\_adaptive\_mutex}, six compaction\ threads) and read-heavy access ({\tt optimize\_filters\_for\_hits}). We prioritized write latency by turning off Write Ahead Log (WAL) compression ({\tt wal\_compression}) and setting a 64 GiB total WAL size. We used this configuration for all subsequent experiments on compressor selection, block size, and Terabyte-scale performance.

\section{Micro-Benchmarks on a 10 GiB Python Corpus}
\label{sec:micro-bench}

We established a single-threaded performance baseline using the 10 GiB corpus of 434,322 Python files. We tested build insertions, single- and multi-key retrieval, accessing keys in their original (sorted) dataset order; for multi-gets, we used batches of 100 keys. We also evaluated a power-law key distribution ($\alpha = -1.5$) to model real-world scenarios where users query similar items concurrently, useful to mimic a best-case {\em warm start} for frequent keys or terms~\cite[\S5.1.2]{IR-book} or Web search~\cite{power2000, power-law-distr}. We averaged energy measurements over five executions.
\Cref{fig:benchmark} presents the results. Upper subplots (grey background) show data throughput (MB/s); lower subplots (green background) show energy efficiency (MB/J). Superior configurations are in the top-left corner of each plot. Marker shapes denote compression algorithms, color-coded by compression level; each series corresponds to block sizes of 4, 16, 64, and 128 KiB, read from right to left, with increasing block size improving compression. Gray lines show the uncompressed baseline. Figures report a strong time--energy correlation, so in the following, we limit our considerations to the upper plots. A clear trade-off emerges: larger block sizes improve compression, do not significantly alter insertion throughput, but slow retrieval, as seen in the horizontal, right-to-left trend for insertion and the vertical, top-down trend for retrieval. Increasing the compression level in \zstd{} and \zlib{} improves insertion but degrades single-get retrieval. \zstd{} outperforms \zlib{} in both metrics, leading us to discard \zlib{} from subsequent experiments. We exclude Snappy too, as it offers only marginally better insertion than \zstd{}, at about double the storage footprint and with worse retrieval.
Hence, we focus on \zstd{}. For the levels 9, 12, and 22, it exhibits similar retrieval but divergent insertion. \zstdlvl{22}'s insertion peaks at 7.8 \mbs; this would require $\approx$29 hours to index the 800 GiB of \Cref{subsec:datasets}. \zstdlvl{12} is faster yet still more then $2\times$ slower in insertion than \zstdlvl{9} for a negligible (0.6\%) single-gets gain and similar compression ratios. For levels 3 and 6, we observe faster insertion, but a slight reduction in compression (up to 2\%) and retrieval (a relative $8.4$\% to $10.9$\% reduction). As for the remaining multi-gets and power-law plots, we observe similar domination trends at increased speed --- computations are over one order of magnitude faster for power-law multi-gets.
As a result of this micro-benchmark analysis, we adopt four large-scale configurations:

\begin{itemize}
  \item \zstdlvl{3} with 64 KiB block, dominating smaller blocks in insertion with competitive retrieval.

  \item \zstdlvl{6} with 4 KiB block, maximizing single-get speed, and with 128 KiB block, yielding better compression than \zstdlvl{3} and similar retrieval.

  \item \zstdlvl{9} with 128 KiB block, dominating in compression \zstdlvl{6} with comparable insertion and retrieval.

\end{itemize}

\begin{figure*}

  \centering
  \begin{subfigure}[t]{.48\linewidth}
    \caption{Uniform single-get}
    \includegraphics[width=\textwidth]{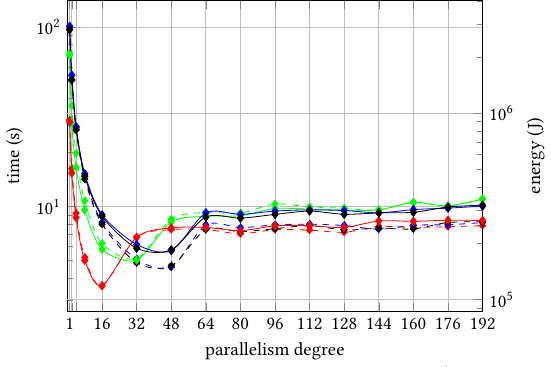}
  \end{subfigure}
  \hfill
  \begin{subfigure}[t]{.48\linewidth}
    \caption{Uniform multi-get}
    \includegraphics[width=\textwidth]{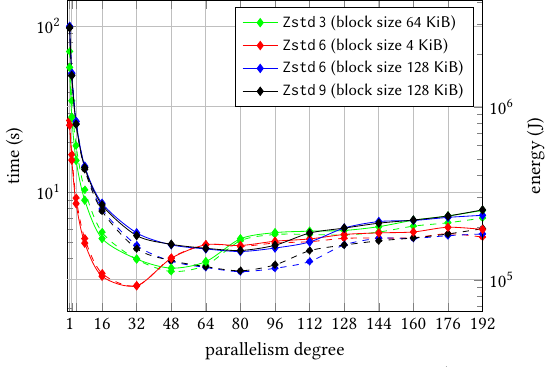}
  \end{subfigure}
  \hfill
  \begin{subfigure}[t]{.48\linewidth}
    \caption{Power-law single-get}
    \includegraphics[width=\textwidth]{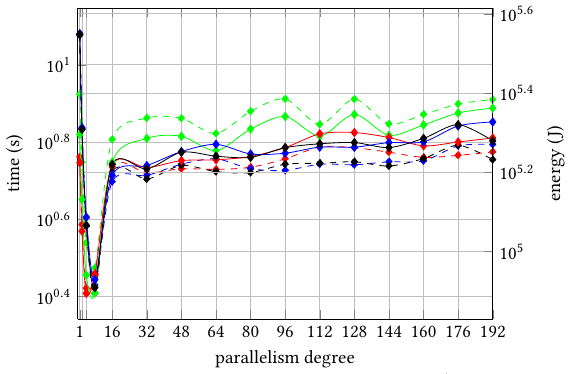}
  \end{subfigure}
  \hfill
  \begin{subfigure}[t]{.48\linewidth}
    \caption{Power-law multi-get}
    \includegraphics[width=\textwidth]{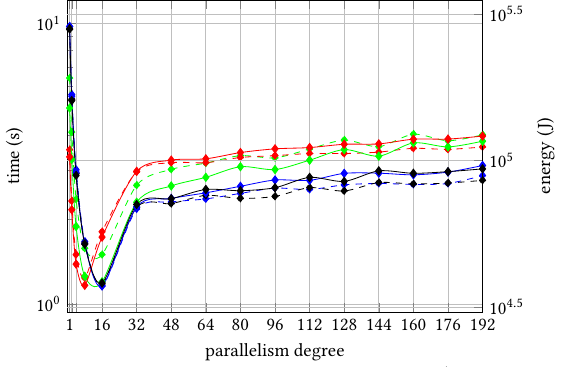}
  \end{subfigure}

  \caption{Time (dashed) and energy (solid) for querying the data in \Cref{tab:boffa_dataset}.}\label{fig:time_energy_boffa}

\end{figure*}
\section{Scaling up to the Large Collections}

\label{sec:scaling}

Though our cache is fully dynamic, herein we evaluate construction (\ref{subsec:build}) and parallel single- and multi-gets (\ref{subsec:get}).

\subsection{Build phase}
\label{subsec:build}

Guided by the considerations in \Cref{sec:micro-bench}, we constructed four separate RocksDB instances, each corresponding to one compressor configuration: \zstdlvl{3} (64 KiB block size), \zstdlvl{6} (4 and 128 KiB), and \zstdlvl{9} (128 KiB). We report our indexing and querying experiments on code data from a Terabyte-scale dataset derived by combining the four 200-GiB language-specific source-code Parquets in \Cref{subsec:datasets}.
The key-sorted order of our Parquet data streamlines efficient indexing in contiguous RocksDB SSTables, since these sequential, predictable accesses enable efficient prefetching and minimize cache misses. The two-level I/O Model~\cite[\S4.2]{pearls_ferragina}\cite{two-level-vitter} indeed clarifies that random accesses yield memory-hierarchy inefficiencies; in contrast, access locality, as in cache-oblivious algorithms~\cite{EfA-cache-oblivious}\cite[\S2]{pearls_ferragina}, mitigates them.

\begin{table}[ht]

    \centering

    \begin{tabular}{llrrrrrr}
        \toprule

 Compr.&Time&Energy&Compr.&Through. \\
 &&& ratio (\%) & (\mbs) \\

        \midrule
        \zstdlvl{3} & 1h 1m 21s & 1086 kJ & 19.05 & 446.67 \\
 (64 KiB) & 3681s & 0.30 kWh  & & \\
        \addlinespace
        \zstdlvl{6} & 2h 24m 2s & 2518 kJ & 19.85 & 190.28 \\
 (4 KiB)    &       8642s & 0.70 kWh  &    &     \\
        \addlinespace
        \zstdlvl{6} & 2h 7m 11s & 2232 kJ & 15.98 & 215.49 \\
 (128 KiB)    &       7631s & 0.62 kWh  &    &     \\
        \addlinespace
        \zstdlvl{9} & 2h 53m 44s & 3047 kJ & 15.49 & 157.75 \\
 (128 KiB)    &       10424s & 0.85 kWh  &    &   \\

        \bottomrule

    \end{tabular}

    \caption{Construction metrics for the data in \Cref{tab:boffa_dataset}.\label{tab:constr}}

\end{table}

\Cref{tab:constr} reports build time, energy, and insertion throughput, with all configurations achieving high throughput (158–447 MB/s). \zstdlvl{9} (128 KiB) outperforms in compression; yet, it is the slowest and most energy-intensive to build. \zstdlvl{3} (64 KiB) had, in turn, a lower compression ratio but was the fastest and most energy-efficient. Lower compression levels optimize speed and energy use, whereas higher levels incur significant costs for marginal compression gains. Our ratios were 15–20\%. We noticed a complex interplay amongst block size and performance: \zstdlvl{6} with 4 KiB blocks was slower than with 128 KiB blocks, suggesting that small-block overhead can outweigh compression gains.
Energy use increased with compression aggressiveness. \zstdlvl{9} (128 KiB) consumed $\approx$0.85 kWh, enough to power an average European electric vehicle for $\approx$4 km~\cite{EU-vehicle} or $\approx3$\% of a US household's daily consumption~\cite{cedar-falls-kwh}. In contrast, \zstdlvl{3} (4 KiB) required only $0.30$ kWh (roughly $1.4$ km for the EV). We note that throughput and energy efficiency are strongly correlated, though not proportional.

\subsection{Multithreaded retrieval}
\label{subsec:get}

We evaluate retrieval using {\em hit} queries, as {\em misses} are handled efficiently by Bloom filters~\cite{bloom_filter}. We measured Parallel throughput to the best serial baseline. To stress random-access performance, we sample one million distinct random keys, potentially forcing each query to decompress a unique block. We thus simulate an inefficient use of the memory hierarchy, aligning with RocksDB's point-lookup evaluation~\cite{rocksdb-evo}. We also test power-law selection ($\alpha = -1.5$), whose repeated keys favor access locality. For multi-gets, we fetch 100 key-value pairs per query in all cases. Thread pools of size~$p$ process queries using {\tt std::future} and {\tt std::async}, following modern C++ concurrency~\cite[\S4.2]{cpp-concurrency-book}.
\Cref{fig:time_energy_boffa} illustrates stream-parallel retrieval for the four languages of \Cref{tab:boffa_dataset}, with thread spanning from 1 to 192. For uniform key distribution and increasing thread count, we note consistently 8.4$\times$ to 30.0$\times$ time performance gains; energy scalability is strongly correlated yet more modest, reflecting the non-energy-proportionality of multicore processors~\cite{non-energy-prop, ieee-access-2025}. Multiple Pareto-optimal configurations exist for a uniform query distribution. \zstdlvl{6} (4 KB) is fastest, optimal at 16 threads for single-gets and at 32 for multi-gets. \zstdlvl{6} (128 KiB) comes second and \zstdlvl{3} offers similar speed at lower thread count. \zstdlvl{9} with 128 KiB is the slowest and requires a higher optimal thread count ($>$80).
For power-law single-gets, behavior inverts: \zstdlvl{6} (4 KiB) is fastest at low parallelism (4 threads). All other configurations require higher parallelism for similar performance. Power-law queries are more rapid than uniform but show limited scalability, with performance plateaus aligning with Amdahl's Law~\cite[\S2.5.4]{parallel_book,amdahl_law}.
For the uniform distribution, speedups range between 8.4-22.0$\times$ (single-gets) and 9.9-30.0$\times$ (multi-gets), thus consistently reducing elapsed time by one to two orders of magnitude. For the power-law case, we achieve 2.2-4.7$\times$ and 3.0-5.4$\times$, respectively. For energy, we obtain similar gain results.
\begin{figure}
  \centering

  \includegraphics[width=.75\linewidth]{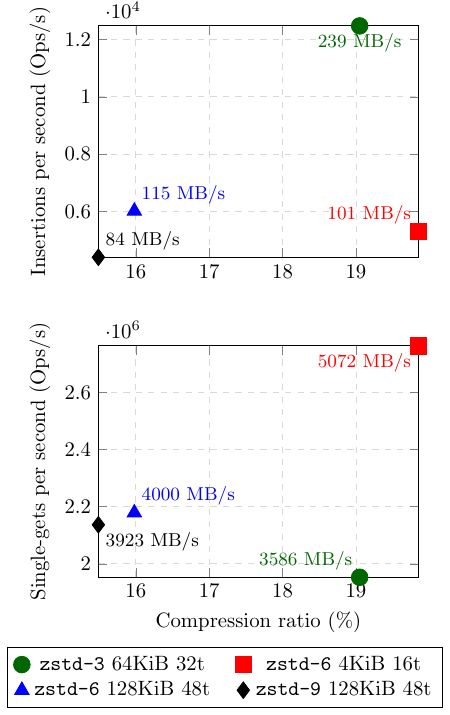}

  \caption{Comparison amongst the best configurations for the four combined datasets in \Cref{tab:boffa_dataset}.\label{fig:scatter}}
\end{figure}
\Cref{fig:scatter} reports space--time Pareto-optimal configurations for build and uniform single-gets derived from \Cref{fig:time_energy_boffa}. The legend identifies color-coded compressor and related level, block size, and optimal thread count for single-gets. In terms of build, \zstdlvl{3} (64 KiB), \zstdlvl{6} (128 KiB), and \zstdlvl{9} (128 KiB) appear on the Pareto frontier, representing distinct speed–compression trade-offs. \zstdlvl{6} (4 KiB), though dominated, requires fewer threads (32) for its optimal configuration, allowing multiple isolated RocksDB instances on each single 192-core node for mirroring, redundancy, or multi-tenancy. The faster retrieval than insertion (one order of magnitude) also aligns with LSM-tree read–write asymmetry~\cite[\S3]{kleppmann_book} (writes require compaction). For single-gets, \zstdlvl{6} (4 KiB) is on the Pareto frontier, as the most time- and energy-performant retrieval algorithm, though space-inefficient.
Our implementation bests \cite{BoffaSWH} in many ways: it is dynamic, achieves similar (15--20\%) compression over larger data, and consistently \gbs retrieval versus their static $\approx$10 \mbs. We reduce {\tt Perf}-profiled energy consumption by one to two orders of magnitude, advancing toward a high-performance solution for dynamic, large-scale environments.

\section{Conclusions and Future Work}\label{sec:conclusions}

Driven by the demand for large-scale, sustainable LLM-era infrastructure, this paper analyzes and experiments with the design of a compressed key-value store for indexing large-scale source code datasets, based on the well-known RocksDB \cite{rocksdb-evo}. Our three-fold evaluation of space, time, and energy performance in RocksDB shows that microbenchmarks and appropriate tuning can yield diverse Pareto-optimal solutions with compelling trade-offs in time-energy retrieval, build efficiency, compression ratios, and thread count trade-offs. Whilst extreme compression levels (e.g., \zstdlvl{3} and \zstdlvl{22}) excel in improving specific computational resources, balanced levels like \zstdlvl{6} and \zstdlvl{9} often provide the most robust solutions.
For future work, we will port our analysis to a distributed cluster by running multiple RocksDB instances on different nodes and partitioning source-code files by programming language. We will also explore integrating LLM-designed keys to improve compression performance further. Finally, we will test the integration of our solution within the open-source Winery infrastructure adopted by SWH.

\section*{Acknowledgments and Availability}
All the computations presented in this paper were performed using the GRICAD infrastructure (\url{https://gricad.univ-grenoble-alpes.fr}), which is supported by Grenoble research communities. The whole codebase to reproduce the experiments is available at \url{https://github.com/ftosoni/green-compressed-storage}.

\bibliographystyle{plain}

\bibliography{\jobname}

\end{document}